\documentstyle[12pt,aasms]{article}

\newcommand{\apl}{\:^{<}_{\sim}\:}

\begin{document}

\title{PRESSURE AT THE ISM-HALO INTERFACE: A REHEATING FREQUENCY DEPENDENCE}

\author{MILAN M. \'{C}IRKOVI\'{C}}
\affil{Astronomical Observatory, Volgina 7, 11000 Belgrade, SERBIA}

\begin{center}

and 

\end{center}
\affil{Astronomy Program, Department of Physics and Astronomy \\
SUNY at Stony Brook, Stony Brook, NY 11794--3800, U.S.A.}

\begin{abstract}
The dependence of the thermal pressure of hot galactic halos on a model
parameter describing the frequency of major reheating episodes during galactic
history is investigated. Pressure on the interface between interstellar
medium and the halo gas is especially interesting, since empirical evidence 
here offers one of the simplest constraints on halo models. It is shown 
that two-phase model of Mo \& Miralda-Escud\'e is sufficiently robust with 
respect to uncertainties in the average interval between reheating. 
\end{abstract}

\keywords{galaxies: evolution---galaxies: halos---galaxies: ISM}

\section{INTRODUCTION}
The existence of very extended galactic halos\footnote{The term ``halo'' will
be preferred over such frequently used terms as ``corona'' or ``envelope'',
although it is implied that they all describe the same physical objects.} has
been suspected for a long 
time (Spitzer 1956) on several grounds. In last two and a half decades, from 
the observations of {\it Copernicus}
and {\it IUE\/} satellites we have learnt that highly ionized gas exists in 
the halo of
Milky Way to much larger scale-heights than it was previously assumed
(for a review, see Savage 1988). So-called high-velocity clouds were 
discovered high
above the plane of the Galaxy, and in other nearby spiral galaxies. 
X-ray observations showed the presence of vast
quantities of hot gas in rich clusters of galaxies (Sarazin 1986), as
well as in smaller compact groups (Saracco \& Ciliegi 1995; Mulchaey et al. 
1996). Cooling flows phenomena
were observed in cluster members (Fabian 1994), as well as in individual 
elliptical galaxies (Nulsen, Stewart, \& Fabian 1984). The extraplanar
optical recombination emission was discovered at large galactocentric distances
in several nearby galaxies
(Donahue, Aldering, \& Stocke 1995; Pildis, Bregman, \& Schombert 1994). 
The depletion of globular cluster gas was interpreted as the consequence 
of ram-pressure stripping, and consequences were drawn thereof (Frank \& 
Gisler 1976; Ninkovi\'c 1985). 

Especially important motivation comes from the QSO absorption line studies. 
It is now clear that, at least at low redshift, a bulk of both metal and
Ly$\alpha$ absorption lines arise in gas which is associated with luminous
galaxies (Sargent, Steidel, \& Boksenberg 1988; Bechtold \& Ellingson 1992; 
Lanzetta et al. 1995; Bowen, Blades, \& Pettini 1996; Chen et al. 1998). 
The idea that the QSO absorbers might arise in halos of normal galaxies
originated with Bahcall \& Spitzer (1969), but the quantitative models
were rare (e.g. Bregman 1981) until recently, when vast accumulated data
on absorber statistics enabled establishing strong empirical constraints
on model parameters (see discussion in Mo \& Morris 1994). Very important 
simple two-phase model of the galactic gaseous halos was put forward 
recently by Mo (1994), and Mo \& Miralda-Escud\'e (1996; hereafter MM96).
Their work shows quite persuasively that formation of the halo structure
as a natural consequence of the galaxy formation and subsequent evolutionary
 processess.

Formation of galaxies implies the collapse of the baryonic
content of intergalactic space and its motion through the non-dissipative dark 
matter halos creating
deep gravitational potential well. A halo of hot gas at virial temperature 
($T_v \equiv \mu V_{\rm cir}^2/2k$, where $\mu$ is the average mass per 
particle, $k$\/ is the Boltzmann constant and $V_{\rm cir}$ is the circular 
velocity, which is assumed to parametrize the total mass of the galaxy) will 
form as the
kinetic energy of the infalling material is thermalized in shocks of this
primordial accretion. As the gas collapses, it is shocked and it
subsequently cools (for an early treatment, see White \& Rees 1978). 
Afterwards, similar situation (although
presumably on smaller scale) arises in course of major galaxy mergers: here we
also have a large mass of gas accreted into the system on a short
timescale and substantial reheating. In the meantime, accretion of ambiental 
IGM is small or entirely negligible. When hot gas cools, it forms the {\it 
cold phase}: photoionized clouds at $T_{\rm cold} \sim 10^4$ K, in pressure 
equilibrium with the hot halo, slowly sinking in the galactic gravitational 
potential. While cold phase may dominate the total gas mass, its filling 
factor is always small, and we shall not deal with it here separately.

The galactocentric radius $r_c$ where the cooling time of the hot gas is the 
{\it cooling radius}. Quantitatively, it is necessary to satisfy
\begin{equation}
\label{minone}
\frac{5\mu k T_v}{2 \Lambda(T_v) \rho(r_c)} = t_M,
\end{equation}
where $\Lambda(T_v)$ is the cooling rate at the virial temperature, and 
$\rho(r_c)$ is the physical density of hot gas at the cooling radius.
Inside it, the conditions of adiabaticity and hydrostatic
equilibrium, lead to the equation of state $P \propto \rho^{5/3}$, where 
$\rho$ is the density of the gas. The ratio of the total gas mass and the 
virial mass is set by the dimensionless parameter $f_g$. Isothermal potential 
of the dark matter creates the density profile 
\begin{equation}
\label{zero}
\rho \propto \left( 1- K \ln \frac{r}{r_c} \right)^{\frac{3}{2}},
\end{equation}
where $K$ is the constant, in the simplest model equal to 0.8. Simple 
calculation shows that pressure at some distance $r$ will behave as $P/k 
\propto (1-K \ln \frac{r}{r_c} )^{5/2}$. 

We see that one of the model parameters is $t_M$, the average interval 
between reheating of the gas, the inverse of what we may call the reheating 
frequency. It is one of the most uncertain parameters. MM96 use 
\begin{equation}
\label{one}
t_M =\frac{t}{1+\Omega_0}
\end{equation}
approximation, where $t$ is the age of the universe, and $\Omega_0$ is the 
total cosmological density parameter. It is used quite successfully by MM96, 
although admitting that a detailed treatment is difficult (due not only to 
the uncertainty in $\Omega_0$, but the power spectrum and physics of structure 
formation as well). While not questioning the goodness of this approximation, 
we would like to show that actual physical quantities resulting from their 
model are not much affected by the choice of $t_M$. 

One of, from the physical point of view, most important predictions is the 
value of the thermal pressure of the hot halo at small galactocentric 
distances, where interface with the ISM occurs. Therefore, in this paper, 
we are investigating the changes in the pressure at this inner boundary of 
the hot halo, when the average time-scale for reheating of the gas is varied. 
This is a part of the more comprehensive study of the boundary conditions of 
the two-phase model, which is currently in progress (\'Cirkovi\'c 1998).

\section{THERMAL PRESSURE AT THE HOT HALO BOUNDARIES}

Formally, thermal pressure in the adiabatic model diverges as $r\rightarrow 
0$. Of course, we know, from both theoretical modelling, and observations of 
Galactic ISM, that the real picture is more complicated. It seems clear that 
assumption of (quasi) hydrostatic equilibrium must be abandoned at some 
fiducial galactocentric radius, comparable to the size of disk in a spiral 
galaxy. This is not necessarily exact ISM-halo interface, since a lot of 
non-equilibrium and still poorly-understood phenomena (``galactic fountain'', 
winds, superbubbles, etc.) will conceivably have the desired effect of 
``softening'' the global pressure profile at $R \apl 50$ kpc. We shall 
neglect these dynamical effects, and following Mo (1994) regard rough size 
of the disk, $R_d =20$ kpc as the inner boundary of the hot halo gas. This is
 justified on a qualitative basis, among other arguments, by the fact that no 
 significant large-scale pressure gradients in the ISM of Milky Way were not 
 observed, including observations toward high-latitude stars at galactocentric
  distance representing significant fraction of our chosen $R_d$.  

It is natural to expect that approximately uniform---when averaged over
individual clouds and intercloud medium---ISM pressure is in equilibrium with
the thermal pressure of the envelopping halo gas for any reasonable choice of
parameters (Spitzer 1978). Therefore, we have calculated pressure $P/k$ at
$R_d$ (for $z=0$, i.e. at the present epoch), as a function of $T_M$, a mean
interval between reheating in major mergers. As is clear from the results
obtained (see Figs. 1 and 2), the model is quite robust to the variations in
$t_M$. For the case of a realistic Population II metallicity ($Z=0.1 \;
 Z_\odot$)
shown in Fig. 1, changing $t_M$ by more than an order of magnitude in the case
of $L_\ast$ galaxy results in change of $P/k$ by a factor of $\sim 2.5$. 
Even if the metallicities are higher (MM96), the same tendency persists, 
as is shown in the plot in Fig. 2 (for the global gas metallicity equal to
 the standard solar, which is unrealistically high).

For comparison, the pressure in the local ISM was estimated by Spitzer 
(1978) as $P/k_{\rm B} = 1700$ K cm$^{-3}$ ($\log P/k = 3.2$), or at large
 sample of measured clouds (Jenkins, Jura, \& Loewenstein 1983)
\begin{equation}
\label{ism}
3.4 \leq \log P/k_{\rm B} \leq 3.8. 
\end{equation}
We note that, especially for higher values of $t_M$ (favored also by the
 equation  (\ref{one}) and other data; see, for example, Keel \& Wu 1995)
  theoretical values of the two-phase model are quite consistent with this
   empirical range. Other theoretical uncertainties, like that in a fraction
    of the virial mass contained in gas (parameter $f_g$ in MM96), seem to
     be much greater than the one considered here.  

\section{DISCUSSION}

Major weaknesses of the theoretical framework this simple, manifest themselves
clearly in the fact that mergers are considered only in terms of reheating
through arising shocks, neglecting the mass input which undoubtedly occurs. On
the other hand, without a detailed understanding of the underlying physics of
merging events, it is difficult to proceed in that direction. It is clear,
though, that if the bulk of the mass deposition occurs at small enough
distances (compared to the virial radius before the merger), hot gas will be
just a transient phase before most of the mass goes into cold, photoionized
clouds. This can be estimated as follows: if $t_M$ is large compared to the
timescale given by $\tau=5 \mu k T_0/2 \Lambda(T_0) \rho(r_0)$ (where 
$T_0=T_v(1-K \ln \frac{r_0}{r_c})$ and $\rho(r_0)$ is given by the equation
 (\ref{zero}), with $r_0$ being the characteristic radius for mass deposition), 
a quasi-stationary state is reached, and the discussion we have followed
applies. 

We have so far discussed only rehetaing in major mergers, which undoubtedly 
occurs from empirical evidence (Keel \& Wu 1995, and references therein).
 Other conceivable reheating mechanisms (starbursts or switching on of a nuclear
source) may be important in the inner part of the halo and should be considered
in the course of a future work. 

We have shown that simple model based on assumptions of adiabaticity and quasi
 hydrostatic equilibrium gives physically acceptable results for the pressure
  at the ISM-halo interface, results which are quite insensitive to the
   assumptions about reheating frequency $1/t_M$. 
Until our theoretical knowledge on merger frequencies (or other reheating
 mechanisms) improves, we are justified in using simple approximations 
 like (\ref{one}). 

\vspace{0.5cm}

The author is happy to acknowledge help of Dr. Hou Jun Mo, who
 kindly provided a cooling code, as well as useful discussion, and 
 Mr. Branislav Nikoli\'c whose hospitality and support were essential for
  completion of this work. Dr Luka \v C. Popovi\'c was, as usual, extremely
   helpful with a friendly advice and technical assistance.

\newpage
Fig. 1. {Thermal pressure of the hot halo at galactocentric distance 
$R_d=20$ kpc at present epoch for $f_g=0.05$ and several characteristic values
of circular velocities of a galaxy, as a function of  average interval between
rehetaing $t_M$. Global halo metallicity is set to $Z=0.1 \; Z_\odot$.
Long-dashed line corresponds to a very massive galaxy ($V_{\rm cir}=300$ km
s$^{-1}$), and shord-dashed line to low-mass galaxy ($V_{\rm cir}=100$ km
s$^{-1}$). The case of a typical $L_\ast$ galactic halo is shown as the solid
line ($V_{\rm cir} \approx 220$ km s$^{-1}$).    We note that $P/k$ stays
approximately the same  with reheating frequency taking all physically 
realistic values.}

Fig. 2. {The same as in Fig. 1, except for the (unrealistic) case of halo
 gas having global metallicity equal to standard $Z_\odot$. Average values for the pressure are slightly lower due to more efficient cooling and consequent depletion of the hot phase.} 

\begin{references}
\reference{} Bahcall, J. N., \& Spitzer, L. Jr. 1969, ApJ, 156, L63

\reference{} Bechtold, J. \& Ellingson, E. 1992, ApJ, 396, 20

\reference{} Bowen, D. V., Blades, J. C., \& Pettini, M. 1996, ApJ, 464, 141

\reference{} Bregman, J. N. 1981, ApJ, 250, 7

\reference{} Chen, H.-W., Lanzetta, K. M., Webb, J. K., \& Barcons, X.  1998,
 ApJ, in press

\reference{} \'Cirkovi\'c, M. M. 1998, in preparation

\reference{} Donahue, M., Aldering, G., \& Stocke, J. T. 1995, ApJ, 450, L45

\reference{} Fabian, A. C. 1994, ARAA, 32, 277

\reference{} Frank, J., \& Gisler, G. 1976, MNRAS, 176, 533

\reference{} Jenkins, E. B., Jura, M., \& Loewenstein, M. 1983, ApJ, 270, 88

\reference{} Keel, W. C., \& Wu, W. 1995, AJ, 110, 129

\reference{} Mo, H. J. 1994, MNRAS, 269, L49

\reference{} Mo, H. J., \& Morris, S. L. 1994, MNRAS, 269, 52 

\reference{} Mo, H. J., \& Miralda-Escud\'{e}, J. 1996, ApJ, 469, 589 (MM96)

\reference{} Mulchaey J. S., Davis, D. S., Mushotzky, R. F., \& Burstein, D.
 1996, ApJ, 456, 80

\reference{} Ninkovi\'c, S. 1985, Ap \& SS, 110, 379

\reference{} Nulsen, P. E. J., Stewart, G. C., \& Fabian, A. C. 1984, MNRAS,
 208, 185

\reference{} Pildis, R. A., Bregman, J. N., \& Schombert, J. M. 1994, ApJ, 
423, 190

\reference{} Saracco, P., \& Ciliegi, P. 1995, A \& A, 301, 348

\reference{} Sargent, W. L. W., Steidel, C. C., \& Boksenberg, A. 1988, ApJ,
 334, 22

\reference{} Savage, B. D. 1988, in QSO Absorption Lines: Probing the Universe, 
eds. Blades, J. C. et al. (Cambridge University Press, Cambridge)

\reference{} Spitzer, L. Jr.  1956, ApJ, 124, 20

\reference{} Spitzer, L. Jr. 1978, ``Physical Processes in the Interstellar
 Medium'' (John Wiley \& Sons, New York)

\reference{} White, S. D. M., \& Rees, M. J. 1978, MNRAS, 183, 341 
\end{references}
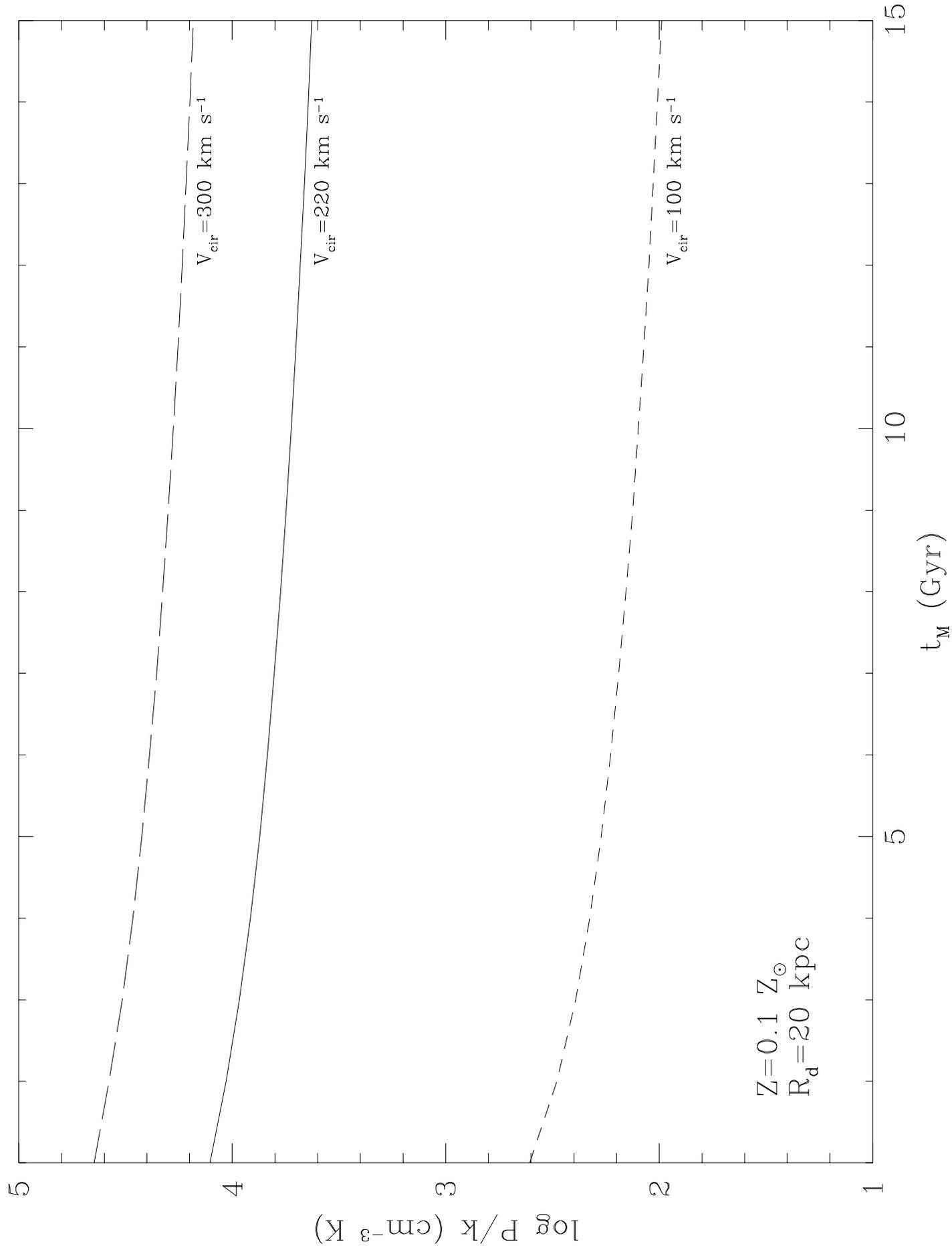
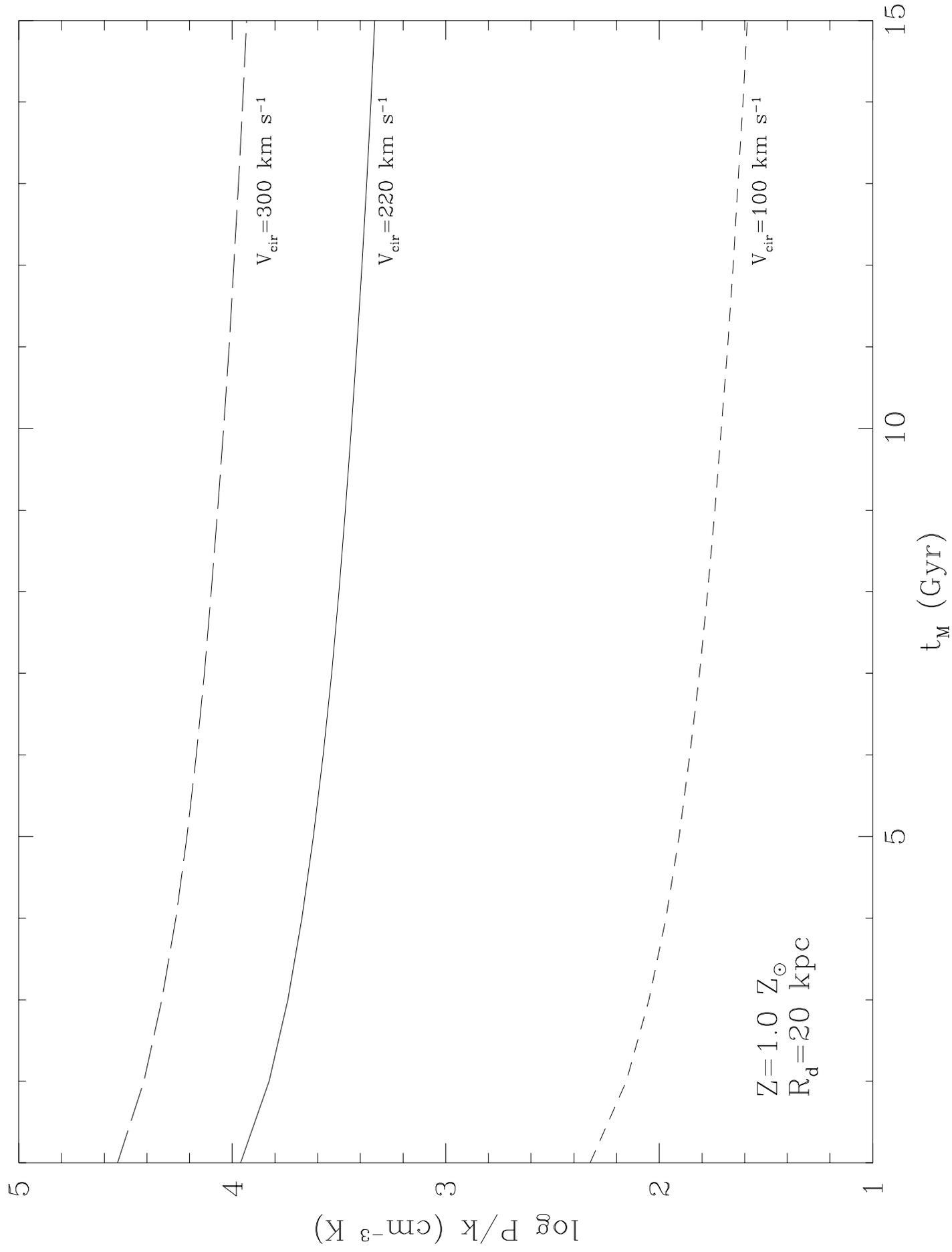
\end{document}